\documentclass [reprint,amsmath,amssymb,aps,prd] {revtex4-2}
\usepackage {CJK}
\usepackage {amsmath}
\usepackage {tikz}
\usepackage {bbm}
\usepackage {bm}
\usepackage {dcolumn}
\usepackage {graphicx}
\usepackage {color}
\usepackage {slashed}

\begin {document}
\title {Schwinger pair creation with the backreaction in 3 + 1 dimensions}
\author {Weitao Liu}
\affiliation {CAS Key Laboratory of Theoretical Physics, Institute of Theoretical Physics, Chinese Academy of Sciences, Beijing 100190, China}
\affiliation {School of Physical Sciences, University of Chinese Academy of Sciences (UCAS), Beijing 100049, China}

\begin{abstract}
	In this work, I analyze the structure of the QED spacetime lattice and review the Schwinger pair creation process from a thermodynamic point of view. This viewpoint enables the dynamical mean-field calculation for the 3 + 1 dimensional Schwinger pair creation with the backreaction. As an example, I demonstrate how to evaluate the pair creation in a finite volume with external electric fields turned on at $t = 0$. The numerical results show how the backreaction responds to the external fields and influences the pair creation.
\end{abstract}

\maketitle

	\section {Introduction}

		To interpret the negative energy levels involved by the free Dirac equation \cite{Dirac1928}, Dirac introduced the hole theory. According to this assumption, when a sea electron is moved to the positive continuum, an electron-positron pair is created. Schwinger \cite {PhysRev.82.664}, Euler and Heisenberg \cite{Heisenberg1936} pointed out that an external strong electric field with $\epsilon = eE_0/m^2 \gtrsim 1$ can induce such a move spontaneously in the vacuum. The phenomenon is called the Schwinger pair creation. It was found that the pair density increases at a constant rate \cite {PhysRevD.82.105026, PhysRevD.78.036008}
		\begin {equation}
			\frac {\dot{N}} {V} = \frac {e^2E_0^2} {4\pi^3} \exp {\left( - \frac {m^2\pi} {eE_0} \right)} \text{.}
			\label {constrate}
		\end {equation}
		The produced pairs create another electromagnetic field, which in turn affect the pair creation. This effect is called the backreaction. Intuitively, at the beginning of the Schwinger pair creation, the backreaction shall be negligible so that the creation rate eq. \eqref {constrate} is exact. With the accumulation of the created pairs, the backreaction shall become significant and eventually stop the pair creation. In spite of the clear physical picture, the quantum mechanical calculation is difficult to approach the final equilibrium since the electromagnetic field is complicated. Progresses have been made lower dimensional problems \cite {Martinez2016, Muschik_2017, PhysRevD.81.085020, Gold2021}, but the 3+1 dimensional case still needs to be investigated due to the unconfinement of the electromagnetic field. \cite {PhysRevD.90.025016} recovered the Schwinger pair creation rate eq. \eqref {constrate} on the lattice without the backreaction,  and include the backreaction in the plasma oscillation. But the late time approaching to the equilibrium is beyoud the validity of the approximation. In this work, the electromagnetic field is regarded as a heat reservior, then the fermionic system is solved in the dynamical mean-field theory. This paper is organized as follows: in Sec. \ref {sec:thermodynamics}, I revisit the Schwinger pair creation in the nonequilibrium thermodynamics and explain why I can analyze this problem in grand canonical ensembles; in Sec. \ref {sec:dmft}, I demonstrate how to organize the dynamical mean field calculation and obtain the charge distribution; in Sec. \ref{sec:rate}, I use the local equilibrium approximation and extract the pair creation rate; in Sec. \ref{sec:conclusion}, I draw a conclusion and point out issues to improve.
	
	\section {The Schwinger pair creation in the nonequilibrium thermodynamics}
		\label {sec:thermodynamics}
		
\begin {figure}
			\resizebox {\columnwidth} {!} {\includegraphics {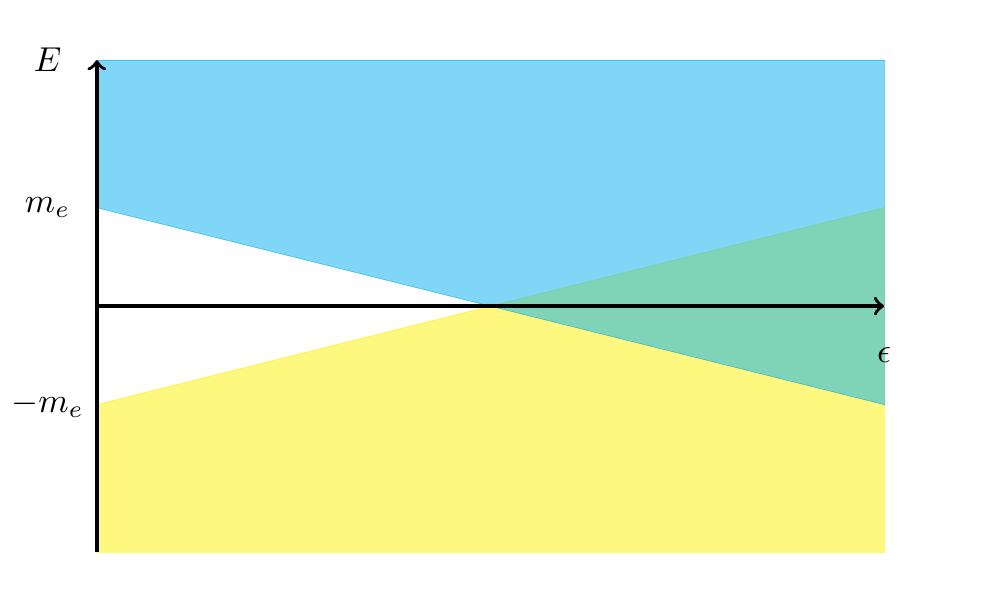}}
			\caption {A schematic illustration of bands mixing in external electric field. Continuous bands are used to replace the discrete levels solved on the lattice.}
			\label {fig:levels}
		\end {figure}

		The Dirac equation leads to a positive and a negative continuum. When an external electric field along the z axis is turned on, the equation has the form
		\begin {equation}
			( - \slashed{\partial} \gamma^0 + m \gamma^0 + E_0z) \Psi = E \Psi \text{.}
			\label {eq:dirac}
		\end {equation}
		Herein and after I use Euclidean Dirac matrices, which satisfy $\{\gamma^\mu, \gamma^\nu\} = 2\delta^{\mu\nu}\mathbbm{1}$. Fig. \ref {fig:levels} illustrates the band structure of eq. \eqref {eq:dirac} with the increasing $\epsilon$. The two bands mix with each other when $\epsilon \sim 1$. Then the sea electrons no longer fill the lower half. Higher half electrons eventually fall down. Projected to the Hilbert space of the free states, this is a process that sea electrons move to the positive continuum. Thus the Schwinger pair creation is a process that the fermionic system spontaneously falls to the new equilibrium. In analogy to the spontaneous emission \cite{doi:10.1119/1.13886}, it is an example of the fluctuation-dissipation theorem \cite{PhysRev.101.1620}, which is accompanied by the energy dissipation from the fermionic system to the electromagnetic field.
	
		This can be seen on the spacetime lattice, where QED is represented by the U(1) K-S Hamiltonian \cite {PhysRevD.11.395}, which can be written as \cite {PhysRevD.15.1128}
		\begin {equation}
			\begin {split}
				H = & \frac {g^2} {4a} \sum_{i,j} A_{ij} l_{ij}^2 + a^3 \sum_{i} \bar{\chi}_{i} (m + V_{i} \gamma^0 ) \chi_{i} \\
				& - \frac {1} {8g^2a} \sum_{i,j,k,l} P_{ijkl} \hat{U}_{ij}\hat{U}_{jk}\hat{U}_{kl} \hat{U}_{li} \\
				& - a^3 \sum_{i,j} \left[ \frac {1} {2a} A_{ij} \bar{\chi}_{i} \vec{\gamma} \cdot \vec{e}_{ij} \hat{U}_{ij} \chi_{j} \right] \text{,}
			\end {split}
			\label {KSHamiltonian}
		\end {equation}
		where the indices $i$ and $j$ denote lattice sites, $A_{ij} = 1$ for $i$ and $j$ are nearest neighbors, otherwise $A_{ij} = 0$, and $P_{ijkl} = 1$ for $i$, $j$, $k$ and $l$ go around the smallest plaquette, otherwise $P_{ijkl} = 0$.
		The operators $l_{ij}$ and $\hat{U}_{ij}$ are related to the electromagnetic field, which satisfy
		\begin {equation}
			\begin {split}
				[l_{ij}, \hat{U}_{ij}] & = - \hat{U}_{ij} \\
				[l_{ij}, \hat{U}_{ji}] & = \hat {U}_{ji} \\
				[l_{ij}^2, l_{ij}] & = 0 \text{.}
			\end {split}
			\label {bosonicoperators}
		\end {equation}
		Due to the commutation relations in eq. \eqref {bosonicoperators}, states of a link can be represented by integers
		\begin {equation}
			\begin {split}
				& \hat{U}_{ij} |n\rangle_{ij} = |n-1\rangle_{ij} \\
				& \hat{U}_{ji} |n\rangle_{ij} = |n+1\rangle_{ij} \\
				& l_{ij} |n\rangle_{ij} = n|n\rangle_{ij} \text{.}
			\end {split}
			\label {bosonicstates}
		\end {equation}
		This representation implies that the electromagnetic field has infinite possible states. On the other hand, a fermionic site $i$ has four components. With the definition $a_\alpha^\dagger = a^{3/2} \chi^\dagger_\alpha$, the operators follow the anti-commutation relation
		\begin {equation}
			\{a_\alpha^\dagger, a_\beta\} = \delta_{\alpha\beta} \text{.} \\
			\label {anticommutation}
		\end {equation}
		This implies that the fermionic component $\alpha$ has only two possible states
		\begin {equation}
			\begin {split}
				& a^\dagger_{\alpha} |0\rangle_\alpha = |1\rangle_\alpha \\
				& a_{\alpha} |1 \rangle_\alpha = |0 \rangle_\alpha \text{.}
			\end {split}
			\label {fermionicstates}
		\end {equation}
		The kinetic operator $a^\dagger_{i} \gamma^0 \vec{\gamma} \cdot \vec{e}_{ij} \hat{U}_{ij} a_{j}$ indicates that when the external potential $V_i$ drives a fermion hop from $j$ to $i$, the electromagnetic field's configuration is changed by the operator $\hat{U}_{ij}$. The corresponding energy change is measured by the electromagnetic energy operator $l_{ij}^2$. Hence the energy from the fermionic system dissipates  into the electromagnetic field.
		
		The infinite number of states makes it difficult to treat the links in the quantum mechanics. However, compared to eq. \eqref {fermionicstates}, eq. \eqref {bosonicstates} implies that the electromagnetic field carries much more microscopic possibilities than the fermionic system. Observing this, I assume that the electromagnetic field provides a heat bath to the fermionic system so that the problem can be studied in the grand canonical ensemble with a temperature $1/\beta$. This temperature is determined by the electromagnetic field. Since the electromagnetic field is a large system, the temperature keeps a constant in the pair creation process. In the numerical calculation, $\beta$ shall be large enough to approximate zero temperature. On the other hand, nonequilibrium problems with external fields in strongly correlated systems were well studied in the framework of the dynamical mean-field theory \cite {PhysRevLett.97.266408, PhysRevB.82.115115, Eckstein2009, PhysRevLett.123.193602}. In these works, nonequilibrium problems within the Hubbard model or the Falicov-Kimball model were mapped onto impurity problems. In this study of Schwinger pair creation, since the electromagnetic field is considered a heat bath, every lattice site is studied as an impurity, then the dynamical mean-field is organized to describe the hopping of electrons between lattice sites. 
	
	\section {The dynamical mean-field theory}
		\label {sec:dmft}	
		
\begin {figure} [h]
			\begin {center}
				\resizebox {0.9\columnwidth} {!} {\includegraphics {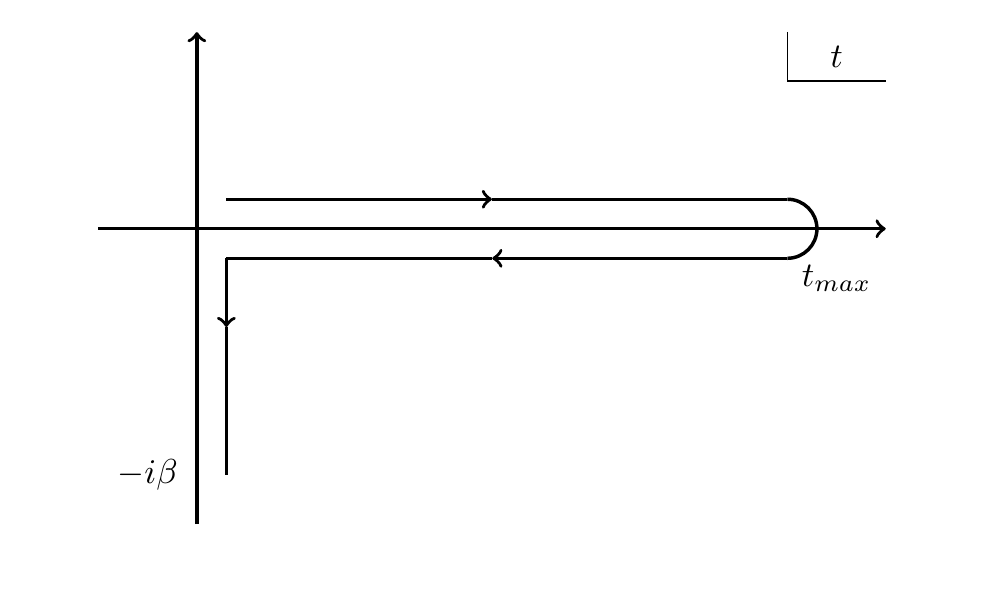}}
				\caption {L-shaped Keldysh contour.}
				\label{KeldyshContour}
			\end {center}
		\end {figure}

		I study every fermionic component as an impurity, its lesser Green's function is defined as
		\begin {equation}
			\begin {split}
				G^{\prec}_{\alpha} (t_1, t_2) = -i \frac {\langle \mathcal{T_C} a_\alpha(t_1) a_\alpha^\dagger(t_2) e^{-iS_\alpha} \rangle} {\langle \mathcal{T_C} e^{-iS_\alpha} \rangle} \text{,} \\ 
			\end {split}
			\label{defgreens}
		\end {equation}
		with the time-ordering operator $\mathcal{T_C}$ orders the operators along the L-shaped Keldysh contour \cite {1963AmJPh..31..309K, KeldyshContour} in Fig. \ref {KeldyshContour}. For two operators $o(t_1)$ and $o(t_2)$,
		\begin {equation}
			\langle \mathcal{T_C} o(t_1) o(t_2) \rangle = \left\{
			\begin {split}
					& -i \langle o(t_1) o(t_2) \rangle \text { for $t_1 \succ t_2$} \\
					& i \langle o(t_2) o(t_1) \rangle \text { for $t_1 \preceq t_2$}
			\end {split} \right. \text{.}
			\label {impuritypartition}			
		\end {equation}
		The action consists of the local part and the dynamical mean-fields part
		\begin {equation}
			\begin {split}
				-iS_\alpha = & {-i \int_\mathcal{C} dt (m_\alpha + V_i(t)) a^\dagger_{\alpha}(t) a_{\alpha}(t)} \\
				& \qquad + i \int_{\mathcal{C}} dt dt' \Lambda_\alpha(t, t') a^\dagger_{\alpha}(t) a_{\alpha}(t') \text{.}
			\end {split}
		\end {equation}
		Because of the $\gamma^0$ matrix, $m_\alpha = m$ for $\alpha$ belonging to the upper two components or $-m$ for the lower two components.
		It is pointed out that the dynamical mean field may come from integrating out other degrees of freedom \cite {Vollhardt2012}. Since a fermionic component is linked to six neighboring components via the $\gamma$ matrices, it consists of the Green's functions of these neighbors. Using different spacings in three dimensions, it reads
		\begin {equation}
			\Lambda_{\alpha} (t, t') = - i \sum_{\beta}|c_{\alpha\beta}|^2 G^\prec_{\beta} (t, t') \text{,}
			\label{dmfs}
		\end {equation}
		with the coefficients $|c_{\alpha\beta}|^2 = (4a_{ij}^2)^{-1}$, where $\alpha \in i$, $\beta \in j$ and $a_{ij}$ is the spacing from $i$ to its neighbor $j$. The Green's function eq. \eqref {defgreens} can be converted to the Grassmann integral \cite {PhysRevD.38.1228} and evaluated by matrix inversions. Hence the system can be solved by iteratively evaluating the Green's function $G_\alpha (t, t')$ for every fermionic component $\alpha$.
		 
\begin {figure}
			\begin {center}
				\resizebox {\columnwidth} {!} {\includegraphics {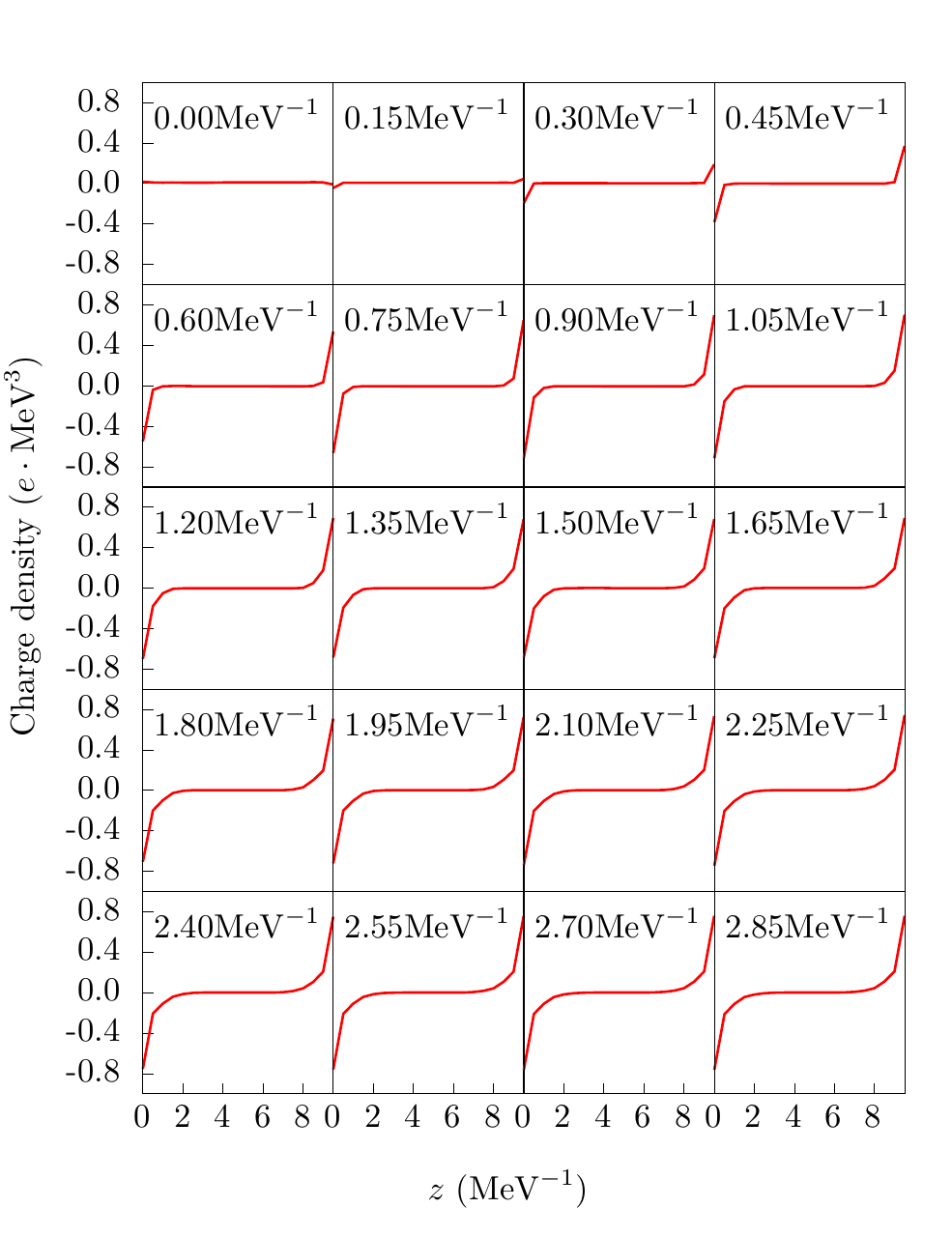}}
				\caption {Charge density evolution in the real time. The results are obtained on a $L_z \times L_y \times L_x = 20 \times 10 \times 10$ lattice with the spacing. The real time spacing $a_t = 0.005 \text{ MeV}^{-1}$ and the imaginary time spacing $a_\tau = 0.04 \text{ MeV}^{-1}$ are small enough to get the convergent result. The choice of lattice spacings $a_z = 0.5 \text{ MeV}^ {-1}$ and $a_x = a_y = 1 \text{ MeV}^ {-1}$ are discussed in Appendix \ref {app:cutoff}}
				\label{ZCharge}
			\end {center}
		\end {figure}

		I consider an external electric field turned on at $t = 0$ in the finite vaccum, where the potential is
		\begin {equation}
			V_i (t) = \left\{
			\begin {split}
				& - i_z a_z \epsilon m \text { for real } t \\
				& 0 \text{ for imaginary } t
			\end {split} \right. \text{.}
		\end {equation}
		Then I solve the Green's function eq. \eqref {defgreens} of every component. The results converge for $\beta > 10 \text{ MeV}^{-1}$. The charge density is extracted from the Green's functions by
		\begin {equation}
			\rho_{i} (t_n) = \left( 2 + i \sum_{\alpha\in i} G^\prec_{\alpha} (t_n, t_n) \right) / a^3 \text{,}
		\end {equation}
		where $a^3 = a_x a_y a_z$ is the volume of a lattice cube. Fig. \ref {ZCharge} illustrates the evolution of the charge density from $t=0$ to $t=2.85\text{ MeV}^{-1}$ with the field $\epsilon = 0.5$. The external field creates electron-positron pairs and separates them to opposite directions. The charge distribution becomes stable around $t = 2 \text{ MeV}^{-1}$, when positive and negative charges concentrate around two boundaries. This line shape is similar to the 1+1 dimensional case in \cite {PhysRevLett.111.201601}. It indicates that the external field is screened by the electron-positron pairs. Hence this result confirms that the backreaction is included in this method.
	
	\section {The pair creation rate}
	\label{sec:rate}
	
		In the last section, I evaluate the Green's functions for every fermionic component. This is insufficient for evaluating more observables which requires $G^\prec_{\alpha\beta}(t,t')$ with $\alpha\neq\beta$. A self energy $\Sigma_\alpha (t, t')$ can be assumed to depict the electromagnetic field's response to the external field. It can be solved by equating the Green's functions
		\begin {equation}
			G^{\prec}_\alpha (t, t') = \mathcal{G}_{\alpha\alpha}^{\prec} (t, t')
			\label {eq:equating} \text{,}
		\end {equation}
		with the definition
		\begin {equation}
			\mathcal {G}_{\alpha\beta}^\prec (t, t') = -i \frac {\langle \mathcal {T}_{\mathcal{C}} a_\alpha(t) a_\beta^\dagger(t') e^{-i\mathcal{S}}\rangle} {\langle \mathcal {T}_{\mathcal{C}} e^{-i\mathcal{S}}\rangle}
			\label {eq:entiregreens} \text{,}
			\end {equation}
		where $\mathcal{S}$ is the action of the fermionic system with the self energy $\Sigma_{\alpha}(t, t')$
		\begin {equation}
			\begin {split}
			- i \mathcal {S} = & - i \int_{\mathcal{C}} dt \sum_{\alpha \in i} (m_\alpha + V_i(t)) a^\dagger_{\alpha}(t) a_{\alpha}(t) \\
				& + i \int_{\mathcal {C}} dt \sum_{\substack {\alpha \in i \\ \beta \in j}} \left[ \frac {A_{ij}} {2a_{ij}}  a^\dagger_{\alpha} (\gamma^0\vec{\gamma} \cdot \vec{e}_{ij})_{\alpha\beta}  a_{\beta} \right] \\
				& + i \int_{\mathcal{C}} dt dt' \sum_\alpha \Sigma_\alpha(t, t') a^\dagger_{\alpha}(t) a_{\alpha}(t') \text{.}
			\end {split}
			\label {eqref:includesigma}
		\end {equation}
		An observable can be evaluated by $ \langle \mathcal {O} (t_1, t_2, \cdots) \rangle = \langle \mathcal {T}_{\mathcal{C}} \mathcal {O} (t_1, t_2, \cdots) e^{-i\mathcal{S}} \rangle / \langle \mathcal{T}_\mathcal{C} e^{-i\mathcal{S}} \rangle$. However, this method calls for very huge amount of computational resources. For example, a system with $L^3 = 1000$ and $2N_r + N_i=1000$, the size of $\mathcal{S}$ matrix is $10^6 \times 10^6$. Solving eq. \eqref {eq:equating} calls for resources exceed the capability of state-of-the-art computers. But for an observable with only one time argument $\langle \mathcal{O}(t) \rangle$, a low-cost alternative is to assume the system stays in a local equilibrium at time $t$. This equilibrium is depicted by the self energy $\Sigma_\alpha (t)$. The action reads
		\begin {equation}
			\begin {split}
			- i \mathcal {S}_\alpha (t) = & - i \int_{\mathcal{I}} d\tau (m_\alpha + V_{i}(t) + \Sigma_\alpha (t)) a^\dagger_{\alpha}(\tau) a_{\alpha}(\tau) \\
				& + i \int_{\mathcal{I}} d\tau d\tau' \lambda_\alpha(t;\tau, \tau') a^\dagger_{\alpha}(\tau) a_{\alpha}(\tau') \text{,}
			\end {split}
		\end {equation}
		with the integration path $\mathcal {I}$ along the imaginary axis from $0$ to $- i \beta$. The dynamical mean field $\lambda$ is now defined as
		\begin {equation}
			\lambda_{\alpha} (t;\tau, \tau') = - i \sum_{\beta}|c_{\alpha\beta}|^2 \mathcal{G}^\prec_{\beta} (t;\tau, \tau') \text{.}
		\end {equation}
		The self energy can be solved by the matching
		\begin {equation}
			G_\alpha^\prec (t, t) = \mathcal {G}_{\alpha}^\prec (t;\tau, \tau) \text{,}
			\label {eq:equating_equilibrium}
		\end {equation}
		where
		\begin {equation}
			\mathcal {G}_{\alpha}^\prec (t; \tau, \tau') = -i \frac{\langle \mathcal {T}_{\mathcal{I}} a_\alpha(\tau) a_\alpha^\dagger(\tau') e^{-i\mathcal{S}_\alpha(t)}\rangle} {\langle \mathcal {T}_{\mathcal{I}} e^{-i\mathcal{S}_\alpha(t)}\rangle} \text{.}
		\end {equation}
		Eq. \eqref {eq:equating_equilibrium} can be solved iteratively. Then an observable $\mathcal {O}(t)$ can be evaluated with
		\begin {equation}
			\langle \mathcal {O} (t) \rangle = \frac {\langle \mathcal {T}_{\mathcal{I}} \mathcal {O}(\tau) e^{-i\mathcal{S}_{\Sigma}(t)} \rangle} {\langle \mathcal {T}_{\mathcal{I}} e^{-i\mathcal{S}_{\Sigma}(t)} \rangle} \text{,}
			\label {eq:localequilibriumexpectation}
		\end {equation}
		with $\mathcal{S}_{\Sigma}$ the fermionic action with the inserted self energies
		\begin {equation}
			\begin {split}
			- i \mathcal {S}_\Sigma (t) = & - i \int_{\mathcal{I}} dt \sum_\alpha (m_\alpha + V_{i}(t) + \Sigma_\alpha (t)) a^\dagger_{\alpha}(\tau) a_{\alpha}(\tau) \\
				& + i \int_{\mathcal {I}} dt \sum_{\substack{\alpha \in i \\ \beta \in j}} \left[ \frac {1} {2} A_{ij} a^\dagger_{\alpha} (\gamma^0\vec{\gamma} \cdot \vec{e}_{ij})_{\alpha\beta}  a_{\beta} \right] \text{.}\\
			\end {split}
		\end {equation}
		To evaluate the total number of pairs, I solve the free Dirac equation on the lattice and get the particle number operator of every state $N_{E_i}$. The total number of pairs equals to the summation of the occupation probabilities of positive energy states or the empty probabilities of negative energy levels. Its expectation value is obtained by using eq. \eqref {eq:localequilibriumexpectation}. The pair density is obtained by dividing the total number of pairs by the volume.
		\begin {figure}
			\begin {center}
				\resizebox {\columnwidth} {!} {\includegraphics {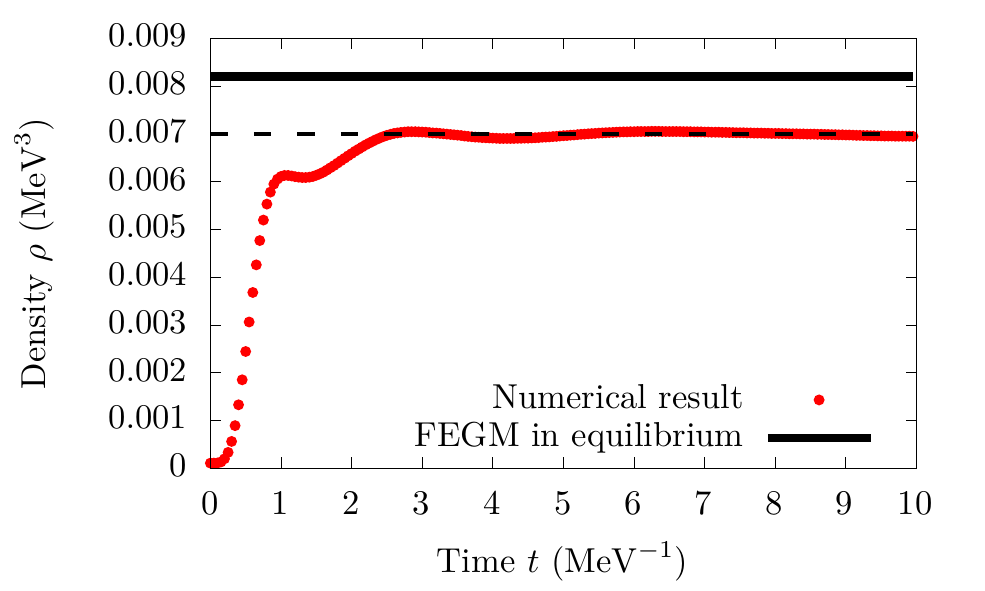}}
				\caption {The evolution of pair density with $\epsilon=0.5$. The dashed line approximates the expectation in the final equilibrium.}
				\label {e05}
			\end {center}
		\end {figure}
		\begin {figure} 
			\begin {center}
				\resizebox {\columnwidth} {!} {\includegraphics {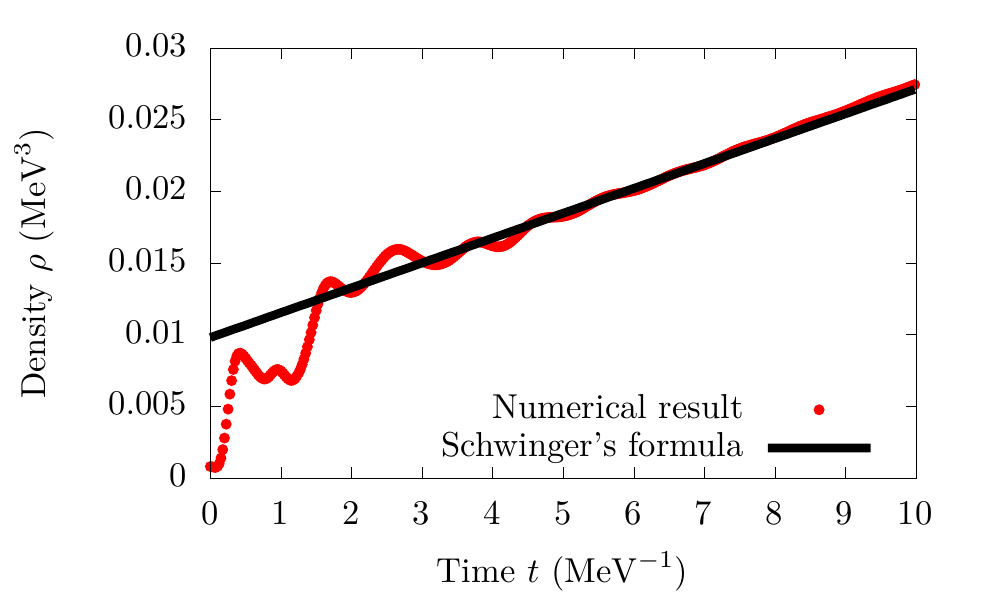}}
				\caption {The evolution of pair density with $\epsilon = 3$. Parameters are the same as in Fig. \ref {ZCharge}}
				\label{e3}
			\end {center}
		\end {figure}
		Fig. \ref {e05} and fig. \ref {e3} illustrate the pair density evolution in a finite space with $\epsilon = 0.5$ and $\epsilon = 3$ respectively. In both cases, the pair density increases drastically once the external field is turned on. This transient enhanced pair creation agrees with the no backreation case in \cite {PhysRevD.90.025016}. But in the case $\epsilon = 0.5$, the pair density does not increase at the rate given by \eqref {constrate} since the system reaches an equilibrium after about $3 \text{  MeV}^{-1}$ with the pair density $0.0070 \text{ MeV}^3$. This is smaller than the free electron gas model, which predicts the pair density $0.0082 \text{ MeV}^3$ in the equilibrium. The dynamical mean-field calculation gives a smaller value due to the backreaction. In the strong electric field $\epsilon = 3$, after the transient enhancement, the pair density increases at a linear rate. This rate agrees with the Schwinger's formula eq. \eqref {constrate}. Hence this method gives an intuitive result. The backreaction is negligible at the beginning in the strong external field.
	
	\section {Conclusions and outlooks}
	\label {sec:conclusion}
	
		In this work, the Schwinger pair creation is a process that the fermionic system falls down to the new equilibrium. Following this idea, I solve the Green's functions in the nonequilibrium dynamical mean-field theory. For strong electric field, at the start time, when the backreaction is negligible, the numerical results agree with \cite {PhysRevD.90.025016}. More importantly, the weak field calculation shows that the method is capable to approach the equilibrium. Thus it is applicable to realistic experiments, such as the upcoming low-energy fully striped heavy ion collisions \cite {GUMBERIDZE2009248, Ma2017}, which are aimed to create strong electric field by the merged ions.
			
		Systematic uncertainties come from two aspects. Firstly, the dynamical mean fields include the Green's functions of the nearest neighbors. Higher order hoppings, such as the cyclings along Wilson loops are not considered. This leads to a cutoff dependence discussed in Appendix \ref {app:cutoff}. To fix this problem, it is necessary to include more neighbors in the dynamical mean fields. Secondly, the local equilibrium approximation loses the information carried by the Green's functions between two different times $G^\prec_\alpha (t, t')$ with $t \neq t'$. To fix this problem, a more reasonable and applicable approximation is needed. The numerical evaluation in this work is essentially iterations of matrix inversions, which consumes the CPU time proportional to $(2N_{\text{r}} + N_{\text{i}})^3$, where $N_\text{r}$ and $N_\text{i}$ are the numbers of time slices along the real and imaginary time axes. For this reason it is difficult to approach the final equilibrium with $\epsilon=3$. A possible solution to this difficulty is to start with a small $t_\text{max}$, since the later time behavior does not impact that in the earlier time, the result can be used to evaluate the Green's function for larger $t_\text{max}$. But many detailed techniques need to be investigated.
	
	\section* {Acknowledgements}
	
		I thank Prof. Ninghua Tong and Prof. Baisong Xie for the helpful discussions. This work is supported by the NSFC and the Deutsche Forschungsgemeinschaft (DFG, German Research Foundation) through the funds provided to the Sino-German Collaborative Research Center TRR110 "Symmetries and the Emergence of Structure in QCD" (NSFC Grant no. 12070131001, DFG Project-ID 196253076-TRR 110), by the NSFC Grant no. 11835015, no. 12047503, and by the Chinese Academy of Sciences (CAS) under Grant no. XDB34030000. The numerical work is done on the HPC Cluster of ITP-CAS. 
	
	\appendix
	\section {Cutoff dependence}
	\label {app:cutoff}
	
		\begin {figure} [t]
			\resizebox {\columnwidth} {!} {\includegraphics {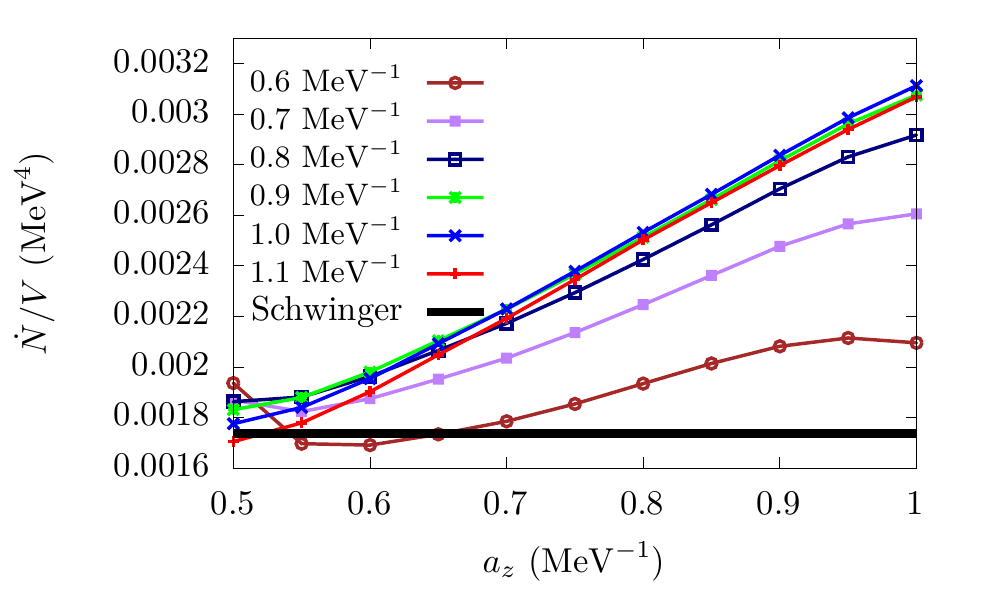}}
			\caption {Density incresing rate dependence on $a_z$ with $\epsilon = 3$.}
			\label {fig:azdep}
		\end {figure}		
		
		In eq. \eqref{dmfs}, for a given fermionic component, the dynamical mean field is approximated by the Green's functions of its neighbors. This approximation includes the hoppings between nearest neighbors. However, higher order hoppings, such as the movement along the Wilson loops, are neglected. Since the hopping coefficient between $i$ and $j$ is $c_{ij} = 1 / a_{ij}$, the choice of lattice spacings shall be large enough so that higher order hoppings are negligible. On the other hand, $a_z$ shall be small enough to resolve the external field in the z- direction. Thus, $a_\perp = a_y = a_x$ shall be kept large enough to suppress higher order hoppings and then small $a_z$ can be used to resolve the external field.
		
		Fig. \ref{fig:azdep} illustrates the cutoff dependence. The density increasing rate $\dot{N}/V$ is determined by a linear fitting to the pair density in the interval $t \in [3\text{ MeV}^{-1}, 10\text{ MeV}^{-1}]$, when the density increases linearly. $a_\perp$ is fixed for every curve. The curves converge for $a_\perp \geqslant 0.9 \text{ MeV}^{-1}$. These curves show that the density increasing rate converges when $a_z$ decreases around $0.5 \text{ MeV}^{-1}$ with a value close to the Schwinger's formula \eqref {constrate}. It shall be noted for $a_\perp = 0.6 \text{ MeV}^{-1}$, the rate rises up when $a_z$ decrease to $0.5 \text{ MeV}^{-1}$. This indicates that the parameter $a_\perp = 0.6 \text{ MeV}^{-1}$ is not large enough to suppress the higher order hoppings when $a_z \leqslant 0.5 \text{ MeV}^{-1}$. Thus I conclude that for a given small $a_z$, one can always use large enough $a_\perp$ to get the converged result.
		
	\bibliography{all}{}

\end {document}